\pdfoutput=1
\PassOptionsToPackage{hyphens}{url}
\documentclass[a4paper]{article}
\usepackage[a4paper,margin=2.5cm]{geometry}
\usepackage[utf8]{inputenc}
\usepackage{subcaption}
\usepackage{graphicx}
\usepackage{pgfplots}
	\pgfplotsset{compat=newest}
	\usepgfplotslibrary{groupplots}
	\usepgfplotslibrary{dateplot}
\usepackage{tikz}
\usepackage{siunitx}

\usepackage{nicefrac}
\usepackage{amsmath}
\usepackage{amsfonts}
\usepackage{amssymb}
\usepackage{mathtools}
\usepackage{bm}
\usepackage{commath}
\usepackage{booktabs}
\usepackage{tabulary}
\usepackage{placeins}
\usepackage{marvosym}

\usepackage[raggedright]{titlesec}

\newcommand\blfootnote[1]{%
  \begingroup
  \renewcommand\thefootnote{}\footnote{#1}%
  \addtocounter{footnote}{-1}%
  \endgroup
}

\usepackage[
	  backend=biber
	, style=authoryear
	, maxcitenames=2
	, maxbibnames=99
	, giveninits=true
	, uniquename=init
	, bibencoding=utf8
	, natbib=true
	, urldate=edtf,date=edtf,seconds=true
	  ]{biblatex}
	\bibliography{refs}
	\AtEveryBibitem{%
		\clearfield{doi}
	}
\usepackage[hidelinks]{hyperref}

\newcommand{\tr}{^\top}
\newcommand{\mtrx}[1]{\bm{\mathrm{#1}}}
\newcommand{\vect}[1]{\bm{#1}}
\newcommand{\mset}[1]{\mathbb{#1}}
\newcommand{\real}{\mathbb{R}}

\DeclareMathOperator{\diag}{diag}
\title{%
	Trajectory Planning and Control for Automatic Docking of ASVs with Full-Scale Experiments
}
\author{%
	Glenn Bitar~\Letter,
	Andreas B.\ Martinsen,
	Anastasios M.\ Lekkas and
	Morten Breivik
}
\date{}

\begin{document}
	\maketitle
	\begin{abstract}
		We propose a method for performing automatic docking of a small autonomous surface vehicle (ASV) by interconnecting an optimization-based trajectory planner with a dynamic positioning (DP) controller for trajectory tracking.
The trajectory planner provides collision-free trajectories by considering a map with static obstacles, and produces feasible trajectories through inclusion of a mathematical model of the ASV and its actuators.
The DP controller tracks the time-parametrized position, velocity and acceleration produced by the trajectory planner using proportional-integral-derivative feedback with velocity and acceleration feed forward.
The method’s performance is tested on a small ASV in confined waters in Trondheim, Norway.
The ASV performs collision-free docking maneuvers with respect to static obstacles when tracking the generated reference trajectories and achieves successful docking.

	\end{abstract}

	\blfootnote{%
		The authors are with the Centre for Autonomous Marine Operations and Systems,
		Department of Engineering Cybernetics,
		Norwegian University of Science and Technology (NTNU),
		O.\ S.\ Bragstads plass 2D,
		NO-7491 Trondheim, Norway
		E-mails: \{glenn.bitar,\allowbreak andreas.b.martinsen,\allowbreak  anastasios.lekkas\}@ntnu.no, morten.breivik@ieee.org.
		The work is supported by the Research Council of Norway through the project number 269116 as well as through the Centres of Excellence funding scheme with project number 223254.

		\vspace{0.5em}
		\noindent
		\textcopyright \, 2020 the authors.
		This work has been accepted to IFAC for publication under a Creative Commons Licence CC-BY-NC-ND.
	}

\section{Introduction} % (fold)
\label{sec:introduction}

Autonomous surface vehicles (ASVs) constitute a topic of significant research and commercial attention and effort.
Motivating factors are economy, flexibility, safety and environmental advantages.
Technology developments in this field are rapid, and the use cases are many, e.g.\ mapping of the ocean floor, military applications such as surveillance, and transportation.
In addition, the relatively low cost of smaller ASVs enables novel concepts, for example autonomous urban passenger ferries that are an alternative to bridges in a city landscape.

To achieve autonomy in transportation operations the following phases must be automated:
\begin{itemize}
    \item Undocking -- moving from the quay in a confined harbor area to open waters,
    \item transit -- crossing a canal or large body of water towards the destination harbor,
    \item docking -- moving from open waters towards the docking position along the quay in a harbor area.
\end{itemize}
Since this paper focuses on the docking phase, we provide a background on automatic docking methods.
The number of reported existing methods is limited in research literature and in commercial applications.
Methods for docking of autonomous underwater vehicles (AUVs) have been introduced by e.g.\ \citet{Rae1992}, \citet{Teo_2015} and \citet{Hong2003}, but they are of limited value for use with surface vessels in a confined harbor area, due to the lack of consideration of nearby obstacles.
\citet{Tran_2012} propose a method for docking of a large ship based on artificial neural networks to control the ship's thrusters, which has shown promising simulation results.
However, this method does not include the harbor layout for collision avoidance.
\citet{Mizuno2015} propose an optimization-based approach taking into account known disturbances.
An optimal nominal path is generated once, and a lower-level model predictive controller (MPC) attempts to follow it.
This method also does not include the harbor layout for collision avoidance, and it is not very realistic to assume known disturbances in such dynamic settings.
Commercial demonstrations of automatic docking have been performed by Wärtsilä\footnote{%
	Wärtsilä press release: \url{https://www.wartsila.com/twentyfour7/innovation/look-ma-no-hands-auto-docking-ferry-successfully-tested-in-norway}.
} and Rolls-Royce\footnote{%
	Rolls-Royce press release: \url{https://www.rolls-royce.com/media/press-releases/2018/03-12-2018-rr-and-finferries-demonstrate-worlds-first-fully-autonomous-ferry.aspx}.
} (now Kongsberg Maritime).
Details about the methods used in these approaches are unavailable to the public.

\begin{figure}[tb]
	\centering
	\includegraphics[width=0.90\linewidth]{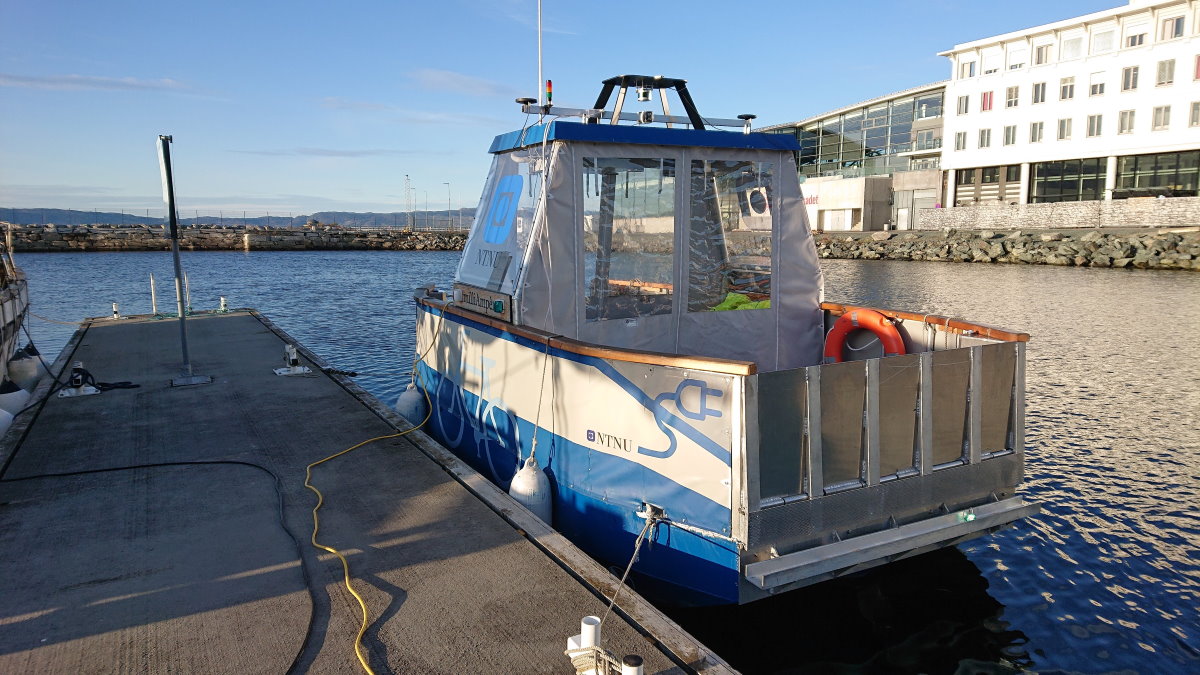}
	\caption{%
		The experimental autonomous urban passenger ferry \emph{milliAmpere}, developed at NTNU, moored near Brattøra in Trondheim, Norway.
	}
	\label{fig:milliampere_moored}
\end{figure}

The docking method from \citep{Martinsen2019docking} is a nonlinear model predictive controller (NMPC) that takes into account vessel dynamics in the form of its dynamic model, as well as collision avoidance by planning trajectories within a convex set, based on the harbor layout.
Advantages of that approach include explicit handling of static obstacles, planning of dynamically feasible trajectories, and flexible behavior shaping via the nonlinear cost function.
The method does not handle moving obstacles or account for external unknown disturbances.
Additionally, due to the non-convex shape of the optimal control problem (OCP), guarantees on run time or feasibility are not provided.
In this paper, we build on \citep{Martinsen2019docking} and add the following contributions:
\begin{itemize}
	\item Instead of running the trajectory planner as an MPC controller by using the inputs directly, the state trajectory is sent to a trajectory-tracking dynamic positioning (DP) controller to account for disturbances and unmodeled dynamics.
 	\item The thruster model is adjusted to improve run times and convexity properties.
 	\item The cost function is adjusted to deal with the wrap-around problem in the heading variable, and to avoid quadratic costs on large position deviations.
 	\item Slack variables are added to deal with feasibility issues that arise when implementing the method in a real-world scenario.
 	\item Although not detailed in this paper, we have implemented an algorithm that dynamically updates the convex set which represents the static obstacles.
 	The set is updated based on the vessel's current position in the map, which allows us to use convex constraints in a non-convex map.
\end{itemize}
By modifying the method from \citep{Martinsen2019docking}, it is shown to produce collision-free and successful maneuvers in full-scale experiments on the experimental autonomous urban passenger ferry \emph{milliAmpere}, seen in \autoref{fig:milliampere_moored}, in Trondheim, Norway.
Although the method is implemented to solve the docking problem on an autonomous urban passenger ferry, this is a generic approach that is suitable also for other use cases and vessel types.

The rest of this paper is structured as follows:
We introduce the experimental platform \emph{milliAmpere} in \autoref{sec:milliampere}.
The trajectory planner used for generating docking trajectories is presented in \autoref{sec:model_predictive_docking_controller}, along with the trajectory-tracking DP controller.
\autoref{sec:experimental_results} presents the experimental results, and we conclude the paper in \autoref{sec:conclusion}.
We present the mathematical models used in the paper in \autoref{apx:models}.

% section introduction (end)

\section{The \emph{milliAmpere} autonomous ferry platform} % (fold)
\label{sec:milliampere}

For the sea trials, we used the experimental autonomous urban passenger ferry \emph{milliAmpere}, depicted in \autoref{fig:milliampere_moored} and with specifications as listed in \autoref{tab:milliampere_specs}.
Developed at the Norwegian University of Science and Technology (NTNU) since 2017, \emph{milliAmpere} has been an experimental platform where many students have contributed with control systems as well as hardware solutions.
A larger version is being designed and built by the research group Autoferry.\footnote{%
	Autoferry website: \url{https://www.ntnu.edu/autoferry}.
}
Small passenger ferries for urban water transport is a novel concept which is being made economically feasible due to increased availability and advances in both sensor systems and autonomous technology.
Such a solution is anticipated to make areas that are separated by waterways more accessible at a lower cost and with less interfering infrastructure than building a bridge.

\bgroup
\def\arraystretch{1.5}
\begin{table}[tb]
	\centering
	\caption{\emph{milliAmpere} specifications.}
	\begin{tabulary}{1.0\linewidth}{LL}
		\toprule
		Dimensions 								& \SI{5}{\meter} by \SI{2.8}{\meter} symmetric footprint \\
		Position and heading reference system 	& Vector VS330 dual GNSS with RTK capabilities \\
		Thrusters 								& Two azimuth thrusters on the center line, \SI{1.8}{\meter} aft and fore of center \\
		Existing control modules				& Trajectory-tracking DP controller and thrust allocation system \\
		\bottomrule
	\end{tabulary}
	\label{tab:milliampere_specs}
\end{table}
\egroup

For simulation purposes, we have used a surge-decoupled three-degree-of-freedom model, along with dynamic models for azimuth angles and propeller speeds of the thrusters.
Separate models are also used for planning and trajectory tracking.
Since we are using three different models in the work described in this paper, we place the model information in \autoref{apx:models} to improve readability.
Parameters and information about the model identification process are available in \citep{Pedersen2019sysid}.

% section milliampere (end)

\section{Trajectory planning and control} % (fold)
\label{sec:model_predictive_docking_controller}

\begin{figure}[tb]
	\centering
	\includegraphics[width=0.98\linewidth]{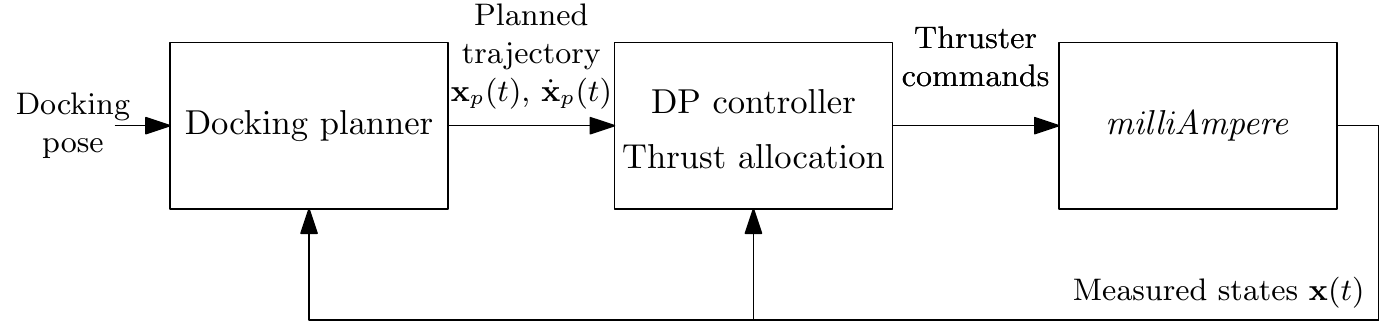}
	\caption{%
		Block diagram of the docking system setup.
		The DP controller and thrust allocation are existing functions on \emph{milliAmpere}.
	}
	\label{fig:block-diagram}
\end{figure}

The trajectory planner is an OCP that takes into account the vessel dynamics via a mathematical model, as well as the harbor layout by including a map as constraints.
The OCP is a modified version of the one developed in \citep{Martinsen2019docking}.
In our case, the OCP runs at a set rate and provides pose, velocity and acceleration trajectories for an existing trajectory-following DP controller, as illustrated in \autoref{fig:block-diagram}.

The OCP is described by the following equations:
\begin{subequations}
	\label{eq:planner-ocp}
	\begin{equation}
		\label{eq:planner-ocp-cost-functional}
		\min_{\vect{x}_{p}(\cdot), \vect{u}_{p}(\cdot), \vect{s}(\cdot)} \int_{t_0}^{t_0+T} \mkern-18mu \left( F(\vect{x}_{p}(t), \vect{u}_{p}(t)) + \vect{k}_s\tr \vect{s}(t) \right) \dif t
	\end{equation}
	subject to
	\begin{align}
		\label{eq:planner-ocp-differential-equations}
		\dot{\vect{x}}_{p}(t) = \vect{f}(\vect{x}_{p}(t), \vect{u}_{p}(t)) & & \forall t \in [t_0, t_0 + T] \\
		\label{eq:planner-ocp-inequality-constraints}
		\vect{h}(\vect{x}_{p}(t), \vect{u}_{p}(t)) - \vect{s}(t) \leq \vect{0} & & \forall t \in [t_0, t_0 + T] \\
		\label{eq:planner-ocp-slack-constraints}
		\vect{s}(t) \geq \vect{0} & & \forall t \in [t_0, t_0 + T] \\
		\label{eq:planner-ocp-initial-conditions}
		\vect{x}_{p}(t_0) = \vect{x}(t_0) & .
	\end{align}
\end{subequations}
The planned states are denoted $\vect{x}_{p} = [\vect{\eta}_{p}\tr, \vect{\nu}_{p}\tr]\tr$,
where $\vect{\eta}_{p} = [x_{p}, y_{p}, \psi_{p}]\tr$ is the Earth-fixed pose, and
$\vect{\nu}_{p} = [u_{p}, v_{p}, r_{p}]\tr$ is the body-fixed velocity vector.
The kinematic relationship between the pose and velocity vectors is detailed in \autoref{apx:models}.
The goal of the OCP is to arrive at the constant state vector
$\vect{x}_{d} = [\vect{\eta}_{d}\tr, \vect{0}\tr_{3}]\tr$ while avoiding collisions, where $\vect{\eta}_{d} = [x_d, y_d, \psi_d]\tr$ is referred to as the \emph{docking pose}.
The vector $\vect{x}(t_0)$ is the vessel's measured state at time $t_0$.
The planning horizon is $T = \SI{120}{\second}$.

The input vector $\vect{u}_{p} = [f_{x1}, f_{y1}, f_{x2}, f_{y2}]\tr$ is used to denote the forces decomposed in surge and sway of \emph{milliAmpere}'s two actuators, where $f_{x1}$ represents a force in surge direction from thruster 1, $f_{y2}$ represents a force in sway direction from thruster 2, etc.
Details on the mapping from this input to control forces are found in \autoref{apx:models}.
The cost functional and constraints are elaborated upon in the following subsections.
The OCP is discretized using direct collocation and solved as a nonlinear program (NLP) with 60 control intervals.

\subsection{Cost functional} % (fold)
\label{sub:cost_functional}

The cost functional \eqref{eq:planner-ocp-cost-functional} operates on the trajectories of the states $\vect{x}_{p}(\cdot)$, inputs $\vect{u}_{p}(\cdot)$ and slack variables $\vect{s}(\cdot)$.
It consists of a cost-to-go function $F(\vect{x}_{p}(t), \vect{u}_{p}(t))$, as well as a cost-to-go on the slack variables $\vect{k}_s\tr \vect{s}(t)$ with the elements of $\vect{k}_s$ having values large enough (\num{1.0e3}) so that the slack variables are active only when the problem otherwise would be infeasible.

The cost-to-go function is
\begin{equation}
	\label{eq:ocp-cost-to-go}
	\begin{gathered}
		F(\vect{x}_{p}(t), \vect{u}_{p}(t)) = \\
		H\!\!\left(
			\begin{bmatrix}
				x_{p}(t) - x_d \\
				y_{p}(t) - y_d
			\end{bmatrix}
		\right) +
		20 \, (1 - \cos(\psi_{p}(t) - \psi_d)) \, +
		10 \, v_p(t)^2 + 10 \, r_p(t)^2 \, +
		\vect{u}_{p}(t)\tr \vect{u}_{p}(t) \, / \, m_{11}^2 \,,
	\end{gathered}
\end{equation}
where the terms are costs on position error, heading error, quadratic sway velocity and yaw rate, and quadratic input, respectively.
The parameter $m_{11}$ is the system inertia in surge, detailed in \autoref{apx:models}.
The terms are scaled so that the cost function becomes dimensionless.
The pseudo-Huber function
\begin{equation}
	H(\vect{a}) = \delta^2 \left(\sqrt{1+\frac{\vect{a}\tr \vect{a}}{\delta^2}}-1\right)
\end{equation}
with $\delta = \SI{10}{\meter}$ provides a quadratic penalty when the quadrature position errors are low and linear when they are high.

The resulting cost functional encourages the planned trajectories to converge to the docking pose $\vect{\eta}_{d}$ with zero velocity, while penalizing sway and yaw rates, as well as the input forces.
Including the docking pose in the cost functional instead of as terminal constraints allows us to use the planner far away from the dock, when the docking pose is outside the reach of the planning horizon $T$.
Additionally, if the operator selects a docking pose that is in violation of the collision constraints, the planner will accept it and find a feasible pose close to the docking pose.

% subsection cost_functional (end)

\subsection{Vessel model} % (fold)
\label{sub:vessel_model}

Equation \eqref{eq:planner-ocp-differential-equations} is a simplified model of the vessel dynamics.
A diagonalized version of the surge-decoupled model in \citep{Pedersen2019sysid} is used, with details found in \autoref{apx:models}.
The kinematic and kinetic models are concatenated to
\begin{equation}
	\dot{\vect{x}}_{p} = \vect{f}(\vect{x}_{p}, \vect{u}_{p}) =
	\begin{bmatrix}
		\mtrx{R}(\psi_{p}) \vect{\nu}_{p} \\
		(\mtrx{S}\,\mtrx{M}_{p})^{-1} ( -\mtrx{C}_{p}(\vect{\nu}_{p}) \vect{\nu}_{p} - \mtrx{D}_{p}(\vect{\nu}_{p}) \vect{\nu}_{p} + \vect{\tau}_{p}(\vect{u}_{p}) )
	\end{bmatrix},
\end{equation}
where the time argument is omitted for notational brevity.
This equation is included as dynamic constraints in the OCP.

% subsection vessel_model (end)

\subsection{Inequality constraints} % (fold)
\label{sub:inequality_constraints}

The inequality constraints \eqref{eq:planner-ocp-inequality-constraints} encode collision avoidance criteria as well as state and input limitations.
These constraints are softened by using slack variables and linear slack costs that keep the optimization problem feasible should disturbances push the vessel outside of these boundaries.

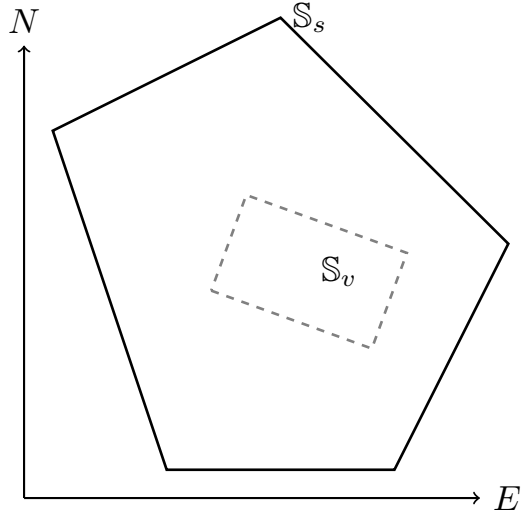
\begin{figure}[tb]
    \centering
    % Define Vessel
\tikzset{
    vessel/.pic = {
        \draw[rounded corners, gray!50, line width=2pt, fill] (25, 15) -- (-25, 15) -- (-25, -15) -- (25, -15) -- cycle;
    }
}

\begin{tikzpicture}[scale = 0.075, transform shape]

    % Rotate and shift from body to NED
    \begin{scope}[shift={(50,40)},rotate=160, scale = 0.5]
        % Vessel center
        \coordinate (body) at (0, 0);
        
        %%%%% Draw vessel %%%%%
        % \pic at (body) {vessel};
        
        % Vectors
        \coordinate (u) at (50, 0);
        \coordinate (v) at (0, -20);
        \coordinate (U) at ($(u) + (v)$);
        
        %%%%% Draw docking margin %%%%%
        \draw[gray, line width=1pt, scale = 1.2, dashed] (25, 15) -- (-25, 15) -- (-25, -15) -- (25, -15) -- cycle;
        
        % Vessel boundary
        \coordinate[scale = 1.2] (boundary) at (20, -9);
        
    \end{scope}
    
    %%%%% Spatial constraints %%%%%
    \begin{scope}[shift={(45,5)}, scale = 0.2]
        \coordinate (spatial) at (0, 400);
        \draw[black, line width=1pt, solid] (-100, 0) -- (100, 0) -- (200, 200) -- (0, 400) -- (-200, 300) -- cycle;
    \end{scope}
    
    %%%%% Set notation %%%%%
    % \node[above right, scale = 12.5] at (boundary) () {$\mset{S}_b$}; 
    \node[right, scale = 17.5] at (body) () {$\mset{S}_v$};
    \node[right, scale = 17.5] at (spatial) () {$\mset{S}_s$};
    
    %%%%% Draw axis %%%%%
    \draw[->, thick] (0,0)--(80,0) node[right, scale = 17.5]{$E$};
    \draw[->, thick] (0,0)--(0,80) node[above, scale = 17.5]{$N$};

\end{tikzpicture}
    \caption{Spatial constraints illustration.}
    \label{fig:spatial_constraints}
\end{figure}

To avoid collisions, we specify a set $\mset{S}_v \subset \real^2$ representing the footprint of the vessel, as well as a permissible convex set $\mset{S}_s \subset \real^2$.
The collision avoidance constraint is to ensure $\mset{S}_v \subset \mset{S}_s$, which can be controlled by checking that the vertices of $\mset{S}_v$ are within $\mset{S}_s$, as illustrated in \autoref{fig:spatial_constraints}.
Since $\mset{S}_s$ is a convex polyhedron, we can describe it as
\begin{equation}
	\label{eq:convex-set-definition}
	\mset{S}_s = \set{\vect{p}\in\real^2~|~ \mtrx{A}_s \vect{p} \leq \vect{b}_s}\,,
\end{equation}
where $\mtrx{A}_s \in \real^{k\times2}$ and $\vect{b}_s \in \real^k$ and $k$ is the number of vertices in the convex set.
This results in the collision avoidance constraint being equivalent to
\begin{equation}
	\label{eq:convex-area}
	\mtrx{A}_s 
	\left(
		\mtrx{R}_2(\psi_{p}(t)) \vect{v} +
		\begin{bmatrix}
			x_{p}(t) \\ y_{p}(t)
		\end{bmatrix}
	\right)
	\leq \vect{b}_s
	\forall~\vect{v} \in \text{Vertex}(\mset{S}_v)\,.
\end{equation}
The rotation matrix $\mtrx{R}_2(\psi_{p}(t))$ is equal to the upper-left $\real^{2\times2}$ of \eqref{eq:apx-rotation-matrix} in \autoref{apx:models}.
The set $\mset{S}_s$ is generated regularly and consists of the eight edges made up of landmasses in the map that are closest to the vessel in order to form a convex set.
Including more edges increases the accuracy of the inequality constraints, but negatively affects run time, and we have found eight to be a good compromise.

The thrusters on \emph{milliAmpere} are each limited in the amount of thrust they are able to produce, so we place limits on the norms of each individual thruster output:
\begin{equation}
	f_{xi}(t)^2 + f_{yi}(t)^2 \leq f_{\text{max}}^2,\quad i \in \{1,2\}\,.
\end{equation}
There are also limits on the states $\vect{x}_{p}$, i.e.
\begin{equation}
	\vect{x}_{lb} \leq \vect{x}_{p}(t) \leq \vect{x}_{ub}\,,
\end{equation}
which ensure that the OCP does not plan trajectories with out-of-bounds velocities.
The limits are only in effect for the velocities in surge and sway ($\pm \SI{1.0}{\meter\per\second}$) and the yaw rate ($\pm \SI{5}{\degree\per\second}$).

As noted, all these constraints are softened with slack variables to ensure feasibility when e.g.\ a disturbance pushes the vessel's state outside the velocity limits or the collision avoidance criterion.
The constraints are gathered in a single vector, giving the inequality constraint vector in \eqref{eq:planner-ocp-inequality-constraints}.

% subsection inequality_constraints (end)

\subsection{Trajectory-tracking DP controller} % (fold)
\label{sub:trajectory_tracking_controller}

The planned state trajectory and its derivative from the solution of \eqref{eq:planner-ocp} are used as reference values for a trajectory-tracking DP controller, which was already implemented on \emph{milliAmpere} before we added the trajectory planner.
There are several reasons for preferring this approach instead of directly using the thruster commands from the solution of \eqref{eq:planner-ocp}:
\begin{itemize}
	\item The planner does not account for drift, disturbances or modeling errors, while the tracking controller does so through feedback.
	\item While the planner is iteration-based with no formal performance guarantees, the tracking controller provides a robust bottom layer that acts also as a safety measure.
	\item The sampling rate of the planner is too low to stabilize the vessel on its own.
\end{itemize}

The tracking controller is based on proportional-integral-derivative (PID) action with feed-forward terms from both velocity and acceleration:
\begin{equation}
	\vect{\tau}_c(t) = \vect{\tau}_{\text{ff}}(t) + \vect{\tau}_{\text{fb}}(t)\,.
\end{equation}
The feed-forward term is
\begin{equation}
	\vect{\tau}_{\text{ff}}(t) = \mtrx{M}_{p} \dot{\vect{\nu}}_{p}(t) + \mtrx{D}_{p}(\vect{\nu}_{p}(t)) \vect{\nu}_{p}(t)\,,
\end{equation}
with details in \autoref{apx:models}.
An issue with this feed-forward term is that it doesn't include coupling Coriolis or damping effects, which may degrade its performance.
This discrepancy is left for the feedback to handle.
The PID feedback is
\begin{equation}
	\label{eq:pid-controller}
	\vect{\tau}_{\text{fb}}(t) = -\mtrx{R}(\psi(t))\tr \cdot
	\left(
		\mtrx{K}_{p} \tilde{\vect{\eta}}(t) +
		\int_0^t \mtrx{K}_i \tilde{\vect{\eta}}(\tau) \dif \tau +
		\mtrx{K}_d \dot{\tilde{\vect{\eta}}}(t)
	\right)\,,
\end{equation}
with $\tilde{\vect{\eta}}(t) = \vect{\eta}(t) - \vect{\eta}_{p}(t)$.
The controller gains are
$\mtrx{K}_{p} = \diag\{
	100,
	100,
	200
\}$,
$\mtrx{K}_i = \diag\{
	10,
	10,
	20
\}$ and
$\mtrx{K}_d = \diag\{
	1000,
	1000,
	1500
\}$ with units that transform the respective elements to force and moment units.
The integrator term in \eqref{eq:pid-controller} has an anti-windup condition, limiting its contribution to $\pm [\SI{150}{\newton}, \SI{150}{\newton}, \SI{200}{\newton\meter}]\tr$.

The control command $\vect{\tau}_c(t)$ is sent to \emph{milliAmpere}'s thrust allocation system, detailed in \citep{Torben2019alloc}, which sends commanded actuator azimuth angles and propeller speeds to the actuators.

% subsection trajectory_tracking_controller (end)

\subsection{Design tradeoffs} % (fold)
\label{sub:design tradeoffs}

In designing the docking system, it has been necessary to compromise between optimality, performance and robustness.
One of the compromises was to separate trajectory planning and motion control.
While it would be possible to run the trajectory planner as an MPC and use the inputs from its solution directly, separation gives several advantages:
\begin{itemize}
	\item A PID controller accounts for steady-state disturbances and corrects for modeling errors, as opposed to the MPC approach.
	\item Using a high-rate feedback controller allows us to run the planner at a low rate, even though the vessel's dynamics are quite fast.
	The planner has run-times between $0.3$ and $0.7$ \si{\second}, which would make it difficult to stabilize the vessel.
	\item Having a trajectory-tracking controller as the bottom control layer makes the docking system more robust to situations where the solver fails to find a feasible solution.
\end{itemize}
Choosing this hybrid structure, where we separate planning from motion control, we have achieved flexibility in the trajectory planner, disturbance rejection through feedback, and robustness to failures in the planning level.

% subsection design tradeoffs (end)

% section model_predictive_docking_controller (end)

\section{Experimental results} % (fold)
\label{sec:experimental_results}

\begin{figure}[tb]
	\centering
	\includegraphics[width=1.00\linewidth,height=8cm,keepaspectratio]{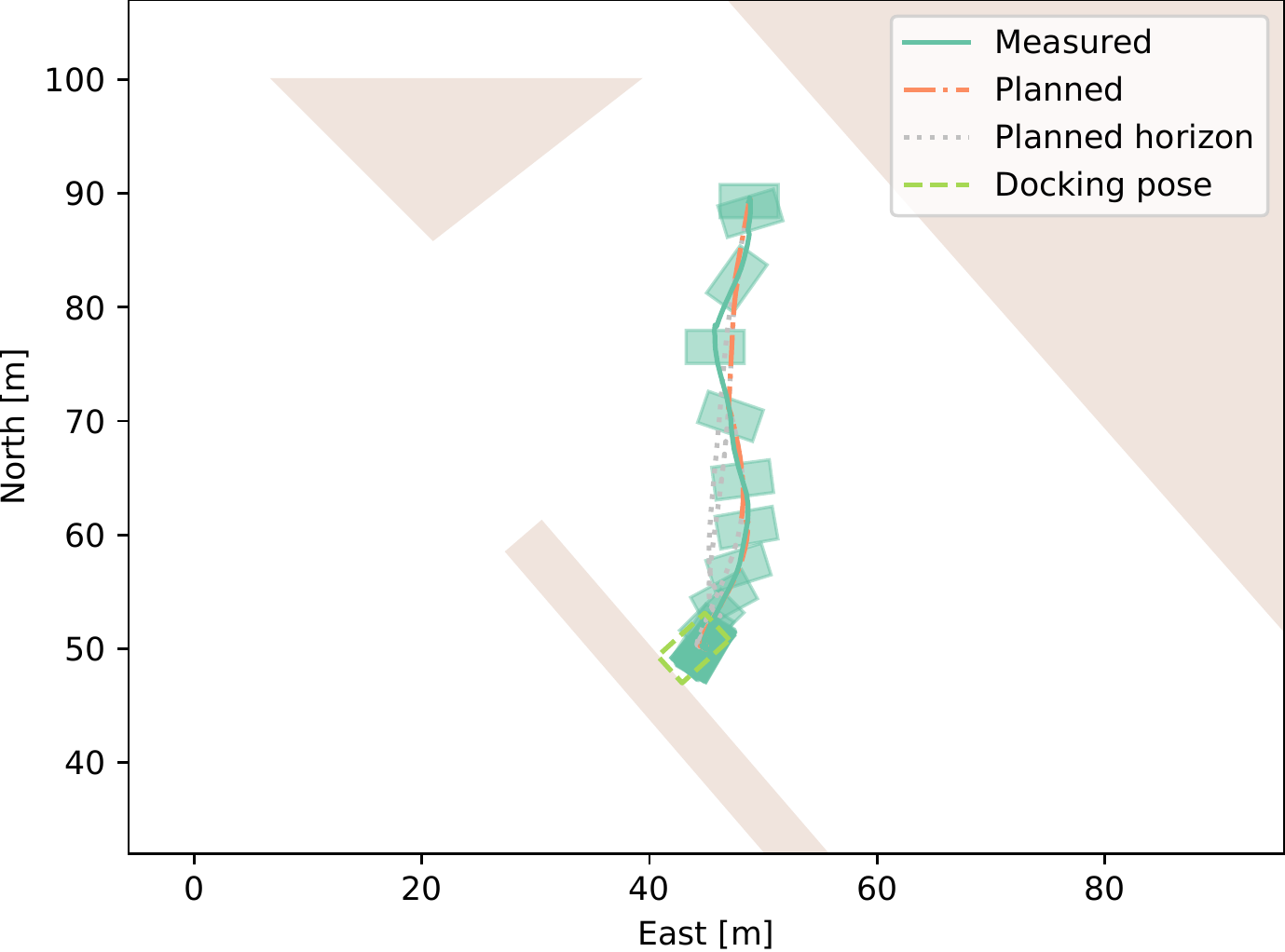}
	\caption{%
		Overview of \emph{milliAmpere}'s trajectory during a docking experiment.
		The vessel's pose is depicted at \SI{5}{\second} intervals with green rectangles.
		The measured position is drawn in solid green, while the active planned reference is in dash-dotted orange.
		The dotted gray lines show the trajectory planner's reference for the entirety of each planning horizon, also after a new solution is calculated.
		The docking pose is marked with a rectangular bright green dashed outline.
	}
	\label{fig:docking_1_xy}
\end{figure}

Experiments were performed with the \emph{milliAmpere} passenger ferry in confined waters in Trondheim, Norway on October 18, 2019.
The weather conditions were calm with winds of \SIrange{2}{3}{\meter\per\second} and rare gusts of \SI{5}{\meter\per\second}.
The vessel is highly susceptible to wind disturbances, due to its large cross-sectional area above water and low underwater profile.
The confined waters protect against waves and currents, however, the shallow depth of \emph{milliAmpere}'s thrusters causes the thrust wake to disturb the hull when operating close to quay, as is the case in the final docking stage.

\begin{figure}[htb]
	\begin{subfigure}[t]{0.45\linewidth}
		\includegraphics[width=1.0\linewidth]{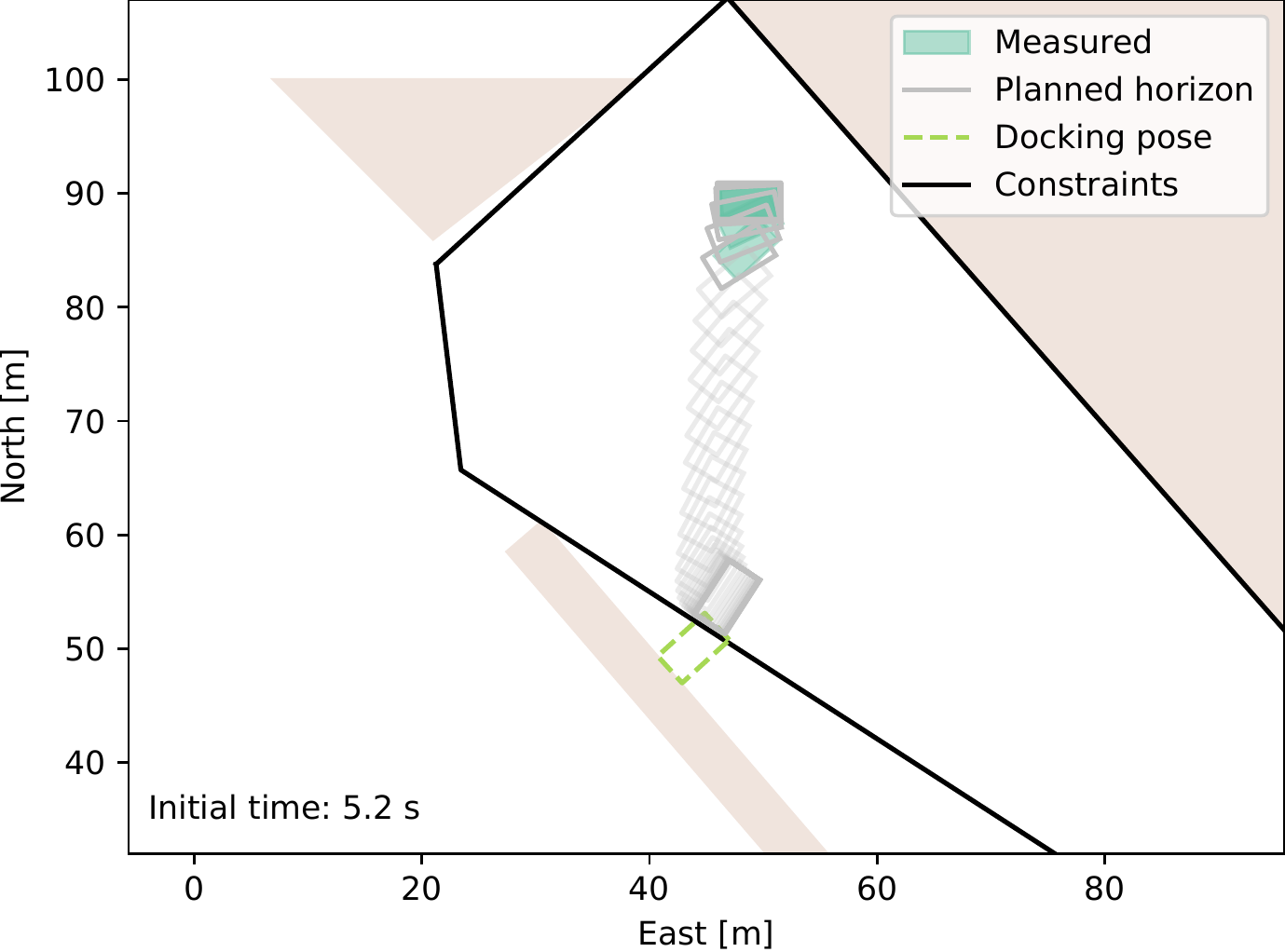}
		\caption{%
			Plots at the first planning step.
		}
		\label{fig:docking_1_plan_00}
	\end{subfigure}
	\begin{subfigure}[t]{0.45\linewidth}
		\includegraphics[width=1.0\linewidth]{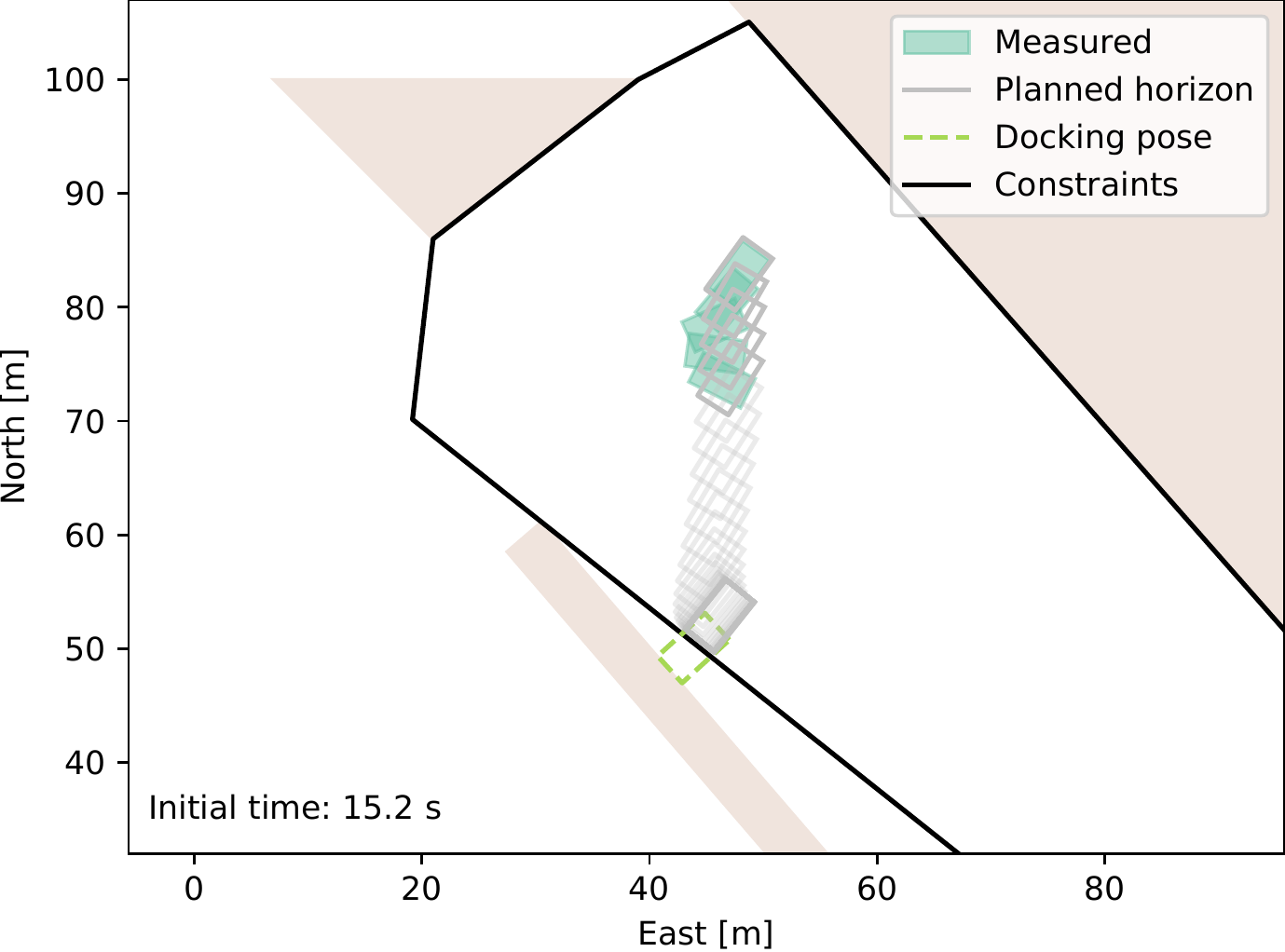}
		\caption{%
			Plots at the second planning step.
			In this step we see that the tracking controller struggles to follow the heading commands.
			We believe this is due to \emph{milliAmpere}'s lack of natural stability in heading, as well as due to poor performance of the DP controller at high velocities.
		}
		\label{fig:docking_1_plan_01}
	\end{subfigure}
	\begin{subfigure}[t]{0.45\linewidth}
		\includegraphics[width=1.0\linewidth]{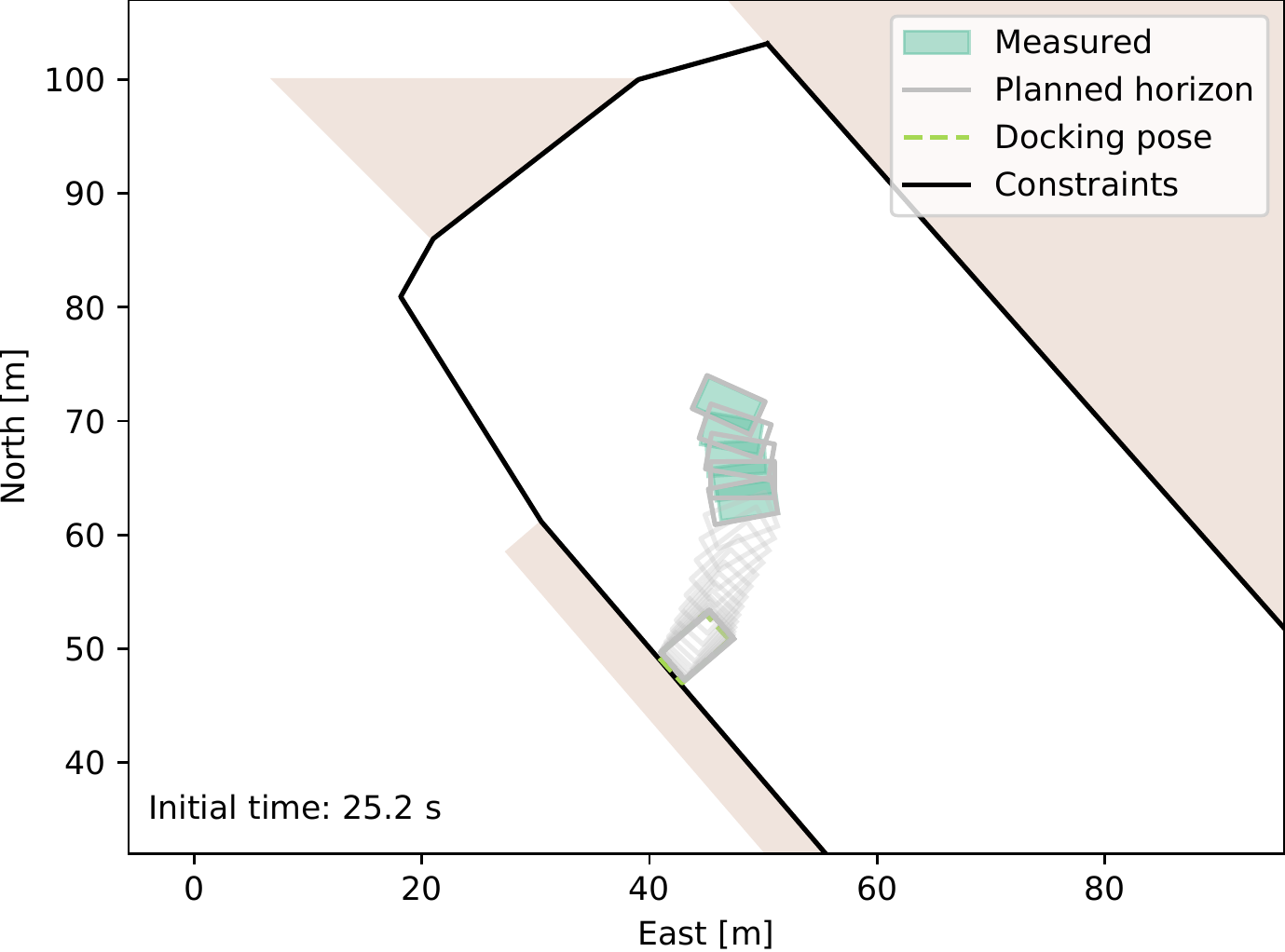}
		\caption{%
			Plots at the third planning step.
			At this slow speed, the tracking controller is able to follow the planned trajectory well.
		}
		\label{fig:docking_1_plan_02}
	\end{subfigure}
	\caption{%
		Planned and measured positions from the first, second and third steps of the planner.
		In \SI{2}{\second} intervals, the plots show the entire planned pose trajectory as gray outlines, and the first \SI{10}{\second} of the measured pose as green rectangles.
		The black solid polyhedron shows the current convex area that represents the spatial constraints from \eqref{eq:convex-area}.
	}
	\label{fig:all_plans}
\end{figure}

To test the docking system, we piloted the ferry to an initial pose around \SI{40}{\meter} away from the docking pose, and activated the docking system once we came to a standstill.
The trajectory planner then calculated state trajectories at a rate of \SI{0.1}{\hertz} towards the docking pose.
A higher rate caused frequent resetting of the error between the planned and measured poses, limiting the effect of the feedback controller \eqref{eq:pid-controller}.
A lower rate would limit the trajectory planner's ability to take into account new information.
Since the trajectory planner calculates a safe trajectory towards the docking pose, a rate of \SI{0.1}{\hertz} is a well-functioning compromise.
Before every run of the planner, an algorithm quickly calculated a new convex area $\mset{S}_s$ based on the vessel's current position, which served as collision-avoidance constraints in the OCP \eqref{eq:convex-set-definition}.
State measurements, the planned trajectory and its derivative were fed to the DP controller at a rate of \SI{10}{\hertz}.
This is sufficient for motion control, since the vessel's dynamics are much slower.

\begin{figure}[tb]
	\centering
	\includegraphics[width=1.0\linewidth,height=12cm,keepaspectratio]{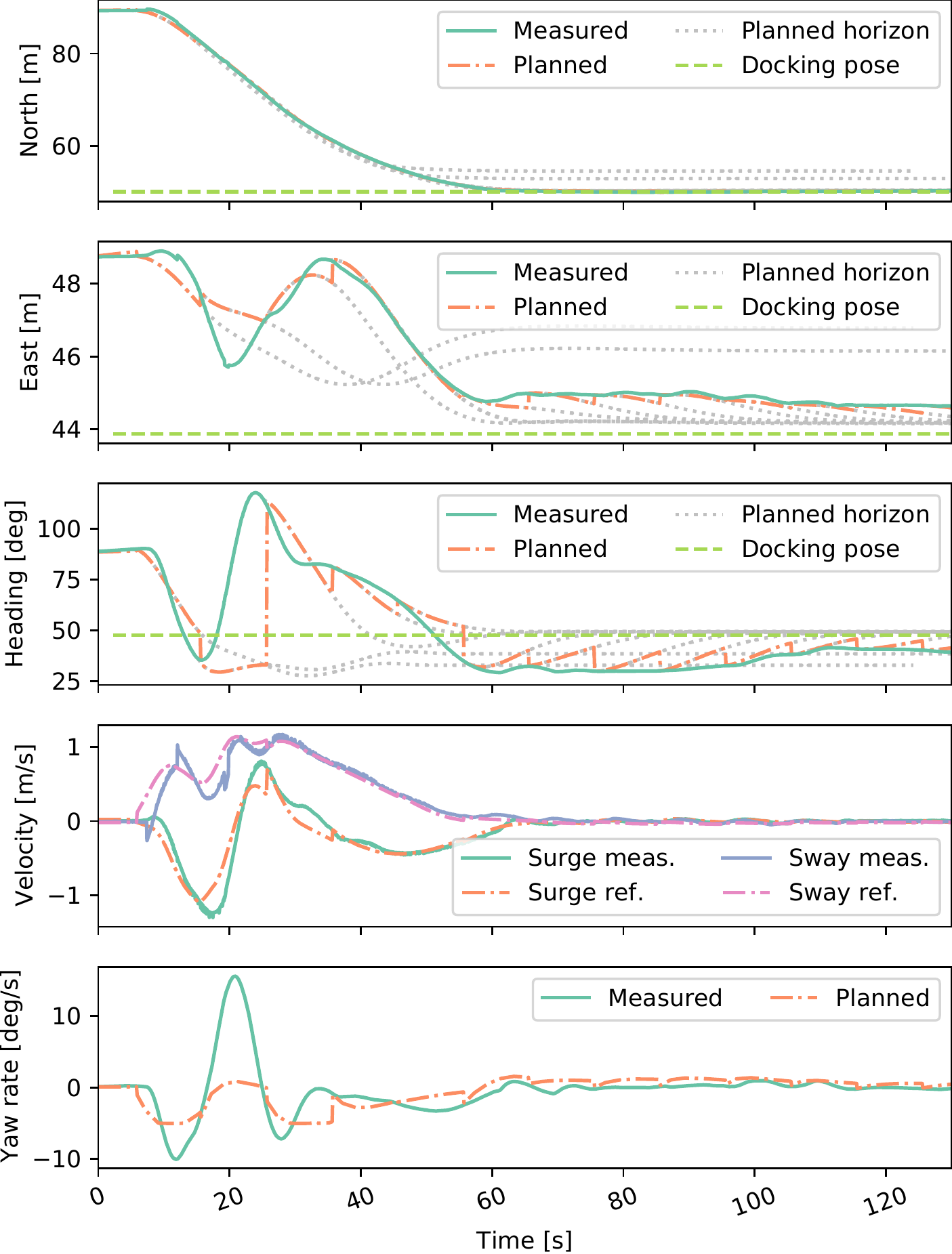}
	\caption{%
		Measured pose and velocity states during the docking experiments, along with reference trajectories and docking pose.
		As in \autoref{fig:docking_1_xy}, we include the full horizon of the planned trajectories in dotted gray lines.
	}
	\label{fig:docking_1_states}
\end{figure}

A bird's eye view of the resulting trajectory is seen in \autoref{fig:docking_1_xy}, with full-state trajectories in \autoref{fig:docking_1_states}.
As is seen in \autoref{fig:docking_1_xy}, \emph{milliAmpere} is able to safely navigate to the docking pose by the help of the docking system.
The trajectory is collision-free and slows down nicely when approaching the quay.
In the course of the experiment there were 13 re-planning steps.
\autoref{fig:all_plans} shows the planned trajectories at the first, second and third steps.
The figure also show the convex areas that the trajectory planner uses for spatial constraints.
Due to how the convex-set algorithm works, the first step does not include the docking pose in its permissible set, so the trajectory planner generates a trajectory towards the edge of its constraints.
The vessel is able to closely follow this trajectory until the second step.
Here we see that the vessel's heading angle is failing to track the planned one.
We believe this is due to \emph{milliAmpere}'s lack of stability in heading, and due to poor tuning of the DP controller, which fails to handle tracking of heading and yaw rate at high speeds.
In the third step, the trajectory planner is able to plan all the way towards the docking pose, and the vessel is able to track the commanded trajectory well, since the speed has decreased.

\autoref{fig:docking_1_states} shows the state trajectories for pose and velocities over time.
It can be seen that the planned trajectories are tracked tightly for the linear positions and velocities.
A notable observation is that the first two plans do not converge to the docking pose, due to the convex area not including the docking pose.
This is corrected as the vessel approaches the harbor.
The heading angle and yaw rate are not converging as well as the linear velocities, especially at high speeds, as seen from the figure.
Additionally, due to the periodic resetting of the planned trajectory to the current vessel state, integration is slow in the DP controller, causing steady-state disturbance rejection to be poor towards the end of the trajectory.

\begin{figure}[tb]
	\centering
	\includegraphics[width=0.88\linewidth,height=3cm,keepaspectratio]{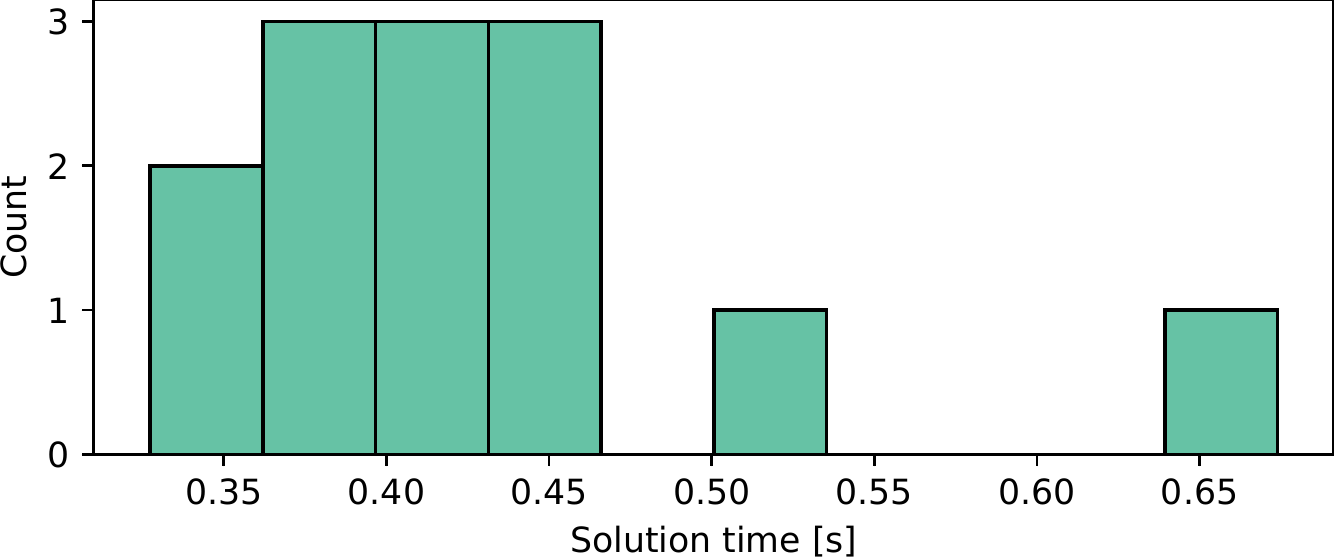}
	\caption{%
		Histogram of solution times of the trajectory planner.
	}
	\label{fig:docking_1_histogram}
\end{figure}

From \autoref{fig:docking_1_histogram}, we see that the solution times of the trajectory planner are in the \SIrange{0.3}{0.7}{\second} range, which is fast enough to be considered real-time feasible when run at a period of \SI{10}{\second}.
These results are repeatable when docking from and to the same pose, and similar results are also seen when docking from other locations.

% section experimental_results (end)

\FloatBarrier

\section{Conclusions and future work} % (fold)
\label{sec:conclusion}

We have demonstrated the capabilities for docking an ASV using an OCP-based trajectory planner in combination with a DP controller.
The solution is tested experimentally in confined waters in Trondheim, Norway, and produces safe maneuvers.
The maneuvers avoid collision with static obstacles and complete the docking phase, ending up in a position adjacent to the dock, ready to moor.
We have shown that the combination of an OCP-based trajectory planner and a tracking controller is suitable for the docking problem.
The method is also general, requiring only a geographic map of sufficient resolution of the harbor environment.
This map may be known a priori, but may also be adjusted with exteroceptive sensors, enabling extensions to the method with camera, lidar and radar systems, e.g.\ using simultaneous localization and mapping techniques.

The experiments have uncovered several possibilities for improvement which are points for future work.
A main conclusion is that although we are able to combine a trajectory planner with an existing tracking controller, the tracking controller must be well-designed and tuned for the combination to function satisfactorily.
The following points are considered as future work:
\begin{itemize}
	\item Improve the tuning of the existing DP controller in order to better track the reference trajectory.
	\item Include coupling effects in the feed-forward term of the DP controller.
	\item Investigate other trajectory-tracking controllers.
	\item Adjust the cost function so that the trajectory planner produces maneuvers that are more consistent with a harbor pilot's experience with docking.
	\item Adjust the trajectory planner to produce more conservative trajectories.
	\item Develop a disturbance estimator that can provide the trajectory planner with valuable information.
\end{itemize}
Future work also includes integrating the docking system in a control structure that handles all the phases of a ferry transport.
The next item in our research is to integrate systems for the undocking and transit phases.
For the undocking phase, the approach presented in this paper is well suited.
For the transit phase, we look to integrate a version of the method from \citep{Bitar2019CamsPipeline}, which can bring the vessel to a location suitable for docking.

% section conclusion (end)

\FloatBarrier

\printbibliography

\appendix

\section{Mathematical vessel models} % (fold)
\label{apx:models}

In this work, we have used three separate models for the \emph{milliAmpere} vessel, respectively for simulation, planning in the OCP, and for trajectory tracking with the DP controller.
All of them are based on the surge-decoupled three-degree-of-freedom model from \citep{Pedersen2019sysid}.
The models use the state vector
\begin{equation}
	\vect{x} =
	\begin{bmatrix}
		\vect{\eta}\tr & \vect{\nu}\tr
	\end{bmatrix}\tr,
\end{equation}
with $\vect{\eta} = [x, y, \psi]\tr \in \real^2 \times S$ being the pose states position north and east of an origin, and heading angle (yaw), respectively.
The velocity vector $\vect{\nu} = [u, v, r]\tr$ contains body-fixed surge velocity, sway velocity and yaw rate, respectively.
The kinematic relationship between the pose and velocity is
\begin{equation}
	\dot{\vect{\eta}} = \mtrx{R}(\psi) \vect{\nu}\,,
\end{equation}
where
\begin{equation}
	\label{eq:apx-rotation-matrix}
	\mtrx{R}(\psi) =
	\begin{bmatrix}
		\cos \psi & -\sin \psi & 0 \\
		\sin \psi & \cos \psi & 0 \\
		0 & 0 & 1
	\end{bmatrix}
\end{equation}
is the kinematic rotation matrix.
The kinetic equations that describe the propagation of $\vect{\nu}$ are different for the three applications.

\subsection{Simulation model} % (fold)
\label{sub:simulation_model}

When we simulated the approach prior to running full-scale experiments, we used the surge-decoupled three-degree-of-freedom model from \citep{Pedersen2019sysid}.
That model has the form
\begin{equation}
	\label{eq:apx-simulation}
	\mtrx{M} \dot{\vect{\nu}} + \mtrx{C}(\vect{\nu}) \vect{\nu} + \mtrx{D}(\vect{\nu}) \vect{\nu} = \vect{\tau}(\vect{\alpha}, \vect{n})\,,
\end{equation}
where $\mtrx{M} \in \real^{3\times3}$ is the positive definite system inertia matrix, $\mtrx{C}(\vect{\nu}) \in \real^{3\times3}$ is the skew symmetric Coriolis and centripetal matrix, and $\mtrx{D}(\vect{\nu}) \in \real^{3\times3}$ is the positive definite damping matrix.
The force vector $\vect{\tau} \in \real^3$ is a function of the thrusters' azimuth angles $\vect{\alpha} = [\alpha_1,\alpha_2]\tr$ and their propeller speeds $\vect{n} = [n_1,n_2]\tr$.
These values are modeled dynamically based on commanded values, with details in \citep{Pedersen2019sysid}.

% subsection simulation_model (end)

\subsection{Planning model} % (fold)
\label{sub:planning_model}

For the dynamic constraints in the OCP, we use a simplified version of \eqref{eq:apx-simulation}:
\begin{equation}
	\label{eq:apx-planning}
	\mtrx{S} \, \mtrx{M}_{p} \dot{\vect{\nu}}_{p} + \mtrx{C}_{p}(\vect{\nu}_{p}) \vect{\nu}_{p} + \mtrx{D}_{p}(\vect{\nu}_{p}) \vect{\nu}_{p} = \vect{\tau}_{p}(\vect{u}_{p})\,,
\end{equation}
where $\mtrx{M}_{p}$, $\mtrx{C}_{p}$ and $\mtrx{D}_{p}$ are diagonalized versions of $\mtrx{M}$, $\mtrx{C}$ and $\mtrx{D}$ from \eqref{eq:apx-simulation}, respectively.
The matrices are
\begin{equation}
	\mtrx{M}_{p} = \diag\{m_{11}, m_{22}, m_{33}\} > 0\,,
\end{equation}
\begin{equation}
	\mtrx{C}_{p}(\vect{\nu}_{p}) =
	\begin{bmatrix}
		0 & 0 & -m_{22} v_{p} \\
		0 & 0 & m_{11} u_{p} \\
		m_{22} v_{p} & -m_{11} u_{p} & 0
	\end{bmatrix}
\end{equation}
and
\begin{equation}
	\mtrx{D}_{p}(\vect{\nu}_{p}) = \diag\{d_{11}(u_{p}), d_{22}(v_{p}), d_{33}(r_{p})\} > 0\,,
\end{equation}
where
\begin{subequations}
	\begin{align}
		d_{11}(u_{p}) &= -X_u - X_{\abs{u}u} \abs{u_{p}} - X_{u^3} u_{p}^2 \\
		d_{22}(v_{p}) &= -Y_v - Y_{\abs{v}v} \abs{v_{p}} - Y_{v^3} v_{p}^2 \\
		d_{33}(r_{p}) &= -N_r - N_{\abs{r}r} \abs{r_{p}}\,.
	\end{align}
\end{subequations}
The coefficient matrix
\begin{equation}
	\mtrx{S} = \diag\{2.5, 2.5, 5.0\}
\end{equation}
is factored into \eqref{eq:apx-planning} to amplify the inertia, making the planned trajectories more sluggish.

The dynamic thruster model from \citep{Pedersen2019sysid} is excluded from the OCP in order to keep the run times down.
Instead, forces from \emph{milliAmpere}'s two thrusters are decomposed in the surge and sway directions, and used directly as inputs:
\begin{equation}
	\vect{u}_{p} = 
	\begin{bmatrix}
		f_{x1} & f_{y1} & f_{x2} & f_{y2}
	\end{bmatrix}\tr,
\end{equation}
where $f_{x1}$ represents a force in surge direction from thruster 1, $f_{y2}$ represents a force in sway direction from thruster 2, etc.
This is mapped to forces and moments in surge, sway and yaw by the function
\begin{equation}
	\vect{\tau}_{p}(\vect{u}_{p}) =
	\begin{bmatrix}
		1 & 0 & 1 & 0 \\
		0 & 1 & 0 & 1 \\
		0 & l_1 & 0 & l_2
	\end{bmatrix}
	\vect{u}_{p}\,.
\end{equation}
The parameters $l_1, l_2 \in \real$ are the distances from the vessel's origin to its thrusters.

% subsection planning_model (end)

\subsection{Tracking controller model} % (fold)
\label{sub:tracking_model}

For the feed-forward terms in the DP controller, we also use a simplified version of the simulation model \eqref{eq:apx-simulation}:
\begin{equation}
	\label{eq:apx-tracking}
	\mtrx{M}_{p} \vect{\nu}_{p} + \mtrx{D}_{p}(\vect{\nu}_{p}) \vect{\nu}_{p} = \vect{\tau}_{\text{ff}}\,.
\end{equation}
The DP controller was originally developed for station keeping, and does not contain the $\mtrx{C}$ matrix.
Otherwise, the matrix values in \eqref{eq:apx-tracking} are equal to those in the planning model \eqref{eq:apx-planning}.

% subsection tracking_model (end)

% section models (end)

\end{document}